\documentstyle[12pt]{article}
\title{Revised Proof of the Uniqueness Theorem for ``No Collapse'' 
Interpretations of Quantum Mechanics}
\author{Jeffrey Bub \\University of Maryland \and Rob Clifton 
\\University of Pittsburgh \and Sheldon Goldstein\thanks{This work 
was supported in part by NSF Grant No. DMS--9504556.}\\Rutgers University}
\date{}

\begin{document}
\maketitle

\begin{abstract}
We show that the Bub-Clifton uniqueness theorem for `no collapse' interpretations 
of quantum mechanics (1996) can be proved without the `weak 
separability' assumption.

\bigskip

\end{abstract}  

\bigskip
\bigskip

Bub and Clifton (1996) proved a uniqueness theorem for `no 
collapse' interpretations of quantum mechanics. The proof was 
repeated, with minor modifications, in (Bub, 1997). The original proof 
involved a `weak separability' assumption (introduced to avoid a 
dimensionality constraint) that required several preliminary 
definitions and considerably complicated the formulation of the theorem. 
Sheldon Goldstein has pointed out that the proof goes through without this 
assumption.

The question at issue is this:  Consider an arbitrary `pure' quantum 
state represented by a ray $e$ on a Hilbert space ${\cal H}$
and the Boolean algebra or lattice, ${\cal B}(R)$, 
generated by 
the eigenspaces of a single 
observable $R$ on ${\cal H}$. The probabilities defined by 
$e$ for the ranges of values of $R$ 
can be represented by a probability measure over the 2-valued 
homomorphisms on   
${\cal B}(R)$. What is the maximal lattice extension 
${\cal D}(e, R)$ of ${\cal B}(R)$, 
generated by eigenspaces of observables other than $R$, on which there exist 
sufficiently many 2-valued homomorphisms so that we can represent the 
probabilities defined by $e$, for ranges of values of $R$ and 
these additional observables, in terms of a measure over the 2-valued 
homomorphisms on ${\cal D}(e, R)$?
The theorem provides an answer to this question on the 
assumption that ${\cal D}(e, R)$ is invariant under 
automorphisms of the lattice ${\cal L(H)}$ of all subspaces of 
Hilbert space that 
preserve the state $e$ and the 
`preferred observable' $R$, without the further assumption of 
`weak separability'.

We sketch a revised proof of the theorem here.

\bigskip
\noindent
\emph{Theorem:} Consider a quantum system $S$ in a (pure) quantum 
state represented by a ray $e$ in an $n$-dimensional 
Hilbert space ${\cal H}$ ($n < \infty)$, and an observable $R$ with $m \leq n$ 
distinct 
eigenspaces $r_{i}$ of ${\cal H}$. 
Let $e_{r_{i}} = (e \vee r_{i}^{\perp}) \wedge 
r_{i}$, $i = 1,2, \ldots, k \leq m$, denote the nonzero projections 
of the ray $e$ onto the eigenspaces $r_{i}$. Then 
${\cal L}_{e_{r_{1}}e_{r_{2}}\ldots e_{r_{k}}} = \{p: e_{r_{i}} \leq p 
\mbox{ or } e_{r_{i}} \leq p^{\perp}, \mbox{ for all }i = 1, \ldots,k\}$ is the 
unique maximal sublattice ${\cal D}(e,R)$ of ${\cal L(H)}$ 
satisfying the following 
three conditions:

(1) \emph{Truth and probability (TP):} ${\cal D}(e,R)$ is an 
ortholattice admitting sufficiently many 2-valued homomorphisms, $h: 
{\cal D}(e,R) \rightarrow \{0,1\}$, to recover all the (single and 
joint) probabilities assigned by the state $e$ to mutually compatible 
sets of elements $\{p_{i}\}_{i \in I}, p_{i}\in {\cal D}(e,R)$, as 
measures on a Kolmogorow probability space $(X, {\cal F}, \mu)$, 
where $X$ is the set of 2-valued homomorphisms on ${\cal D}(e,R)$, 
${\cal F}$ is a field of subsets of $X$, and 
$$
\mu(\{h:h(p_{i}) = 1,\mbox{ for all }i \in I\}) = tr(e\prod_{i \in 
I}p_{i})
$$

(2) \emph{$R$-preferred ($R$-PREF):} the eigenspaces $r_{i}$ of $R$ 
belong to ${\cal D}(e,R)$.

(3) \emph{$e,R$-definability (DEF):} for any $e \in {\cal H}$ and 
observable $R$ of $S$ defined on ${\cal H}$, ${\cal D}(e,R)$ is 
invariant under lattice homorphisms that preserve $e$ and $R$.

\bigskip
\noindent
\emph{Proof:} The strategy of the proof is to show, firstly, that if 
$p \in {\cal D}(e,R)$, then for any $e_{r_{i}}, i = 1,\ldots,k$, 
either $e_{r_{i}} \leq p \mbox{ or } e_{r_{i}} \leq p^{\perp}$, and 
secondly, that the sublattice ${\cal L}_{e_{r_{1}}e_{r_{2}}\ldots e_{r_{k}}} 
= \{p: e_{r_{i}} \leq p 
\mbox{ or } e_{r_{i}} \leq p^{\perp}, i = 1, \ldots,k\}$ satisfies 
the conditions of the theorem. Maximality then requires that ${\cal 
D}(e,R) = {\cal L}_{e_{r_{1}}e_{r_{2}}\ldots e_{r_{k}}}$.

Consider an arbitrary subspace $p$ of ${\cal H}$, and suppose $p \in 
{\cal D}(e,R)$. Clearly, if $r_{i}$ is included in $p$ or in 
$p^{\perp}$, for any $i$, then $e_{r_{i}} \leq p 
\mbox{ or } e_{r_{i}} \leq p^{\perp}$.

Suppose this is not the case. That is, suppose some $r_{i}$ is neither 
included in $p$ nor in $p^{\perp}$. Then either the intersection of 
$r_{i}$ with $p$ or the intersection of $r_{i}$ with $p^{\perp}$ is 
nonzero. (For, by the conditions TP and R-PREF, there must be a 2-valued 
homomorphism mapping $r_{i}$ onto 1, and this homomorphism would have 
to map both $p$ and $p^{\perp}$ onto 0 if $r_{i} \wedge p = r_{i} 
\wedge 
p^{\perp} = 0$, which contradicts TP.) Assume that the intersection 
of $r_{i}$ with $p$ is nonzero (a similar argument applies if we assume 
that the intersection of $r_{i}$ with $p^{\perp}$ is nonzero). Call this 
intersection $b$.

If $p$ belongs to ${\cal D}(e,R)$, then $b = r_{i} \wedge p$ also 
belongs to ${\cal D}(e,R)$ (by the lattice closure assumption of TP, 
since $r_{i}$ belongs to ${\cal D}(e,R)$ by R-PREF). Now, either $b$ is 
orthogonal to $e_{r_{i}}$ in $r_{i}$, or $b$ is skew to 
$e_{r_{i}}$ in $r_{i}$. We shall show that in either case 
$e_{r_{i}}$
must belong to ${\cal D}(e,R)$. It follows that 
$e_{r_{i}} \leq p 
\mbox{ or } e_{r_{i}} \leq p^{\perp}$, because if $e_{r_{i}} \not 
\leq p$ and  $e_{r_{i}} \not \leq p^{\perp}$, then---since 
$e_{r_{i}}$ is a ray---$e_{r_{i}}$ must be skew to both $p$ and 
$p^{\perp}$; that is, $e_{r_{i}} \wedge p = 0$ and $e_{r_{i}} \wedge 
p^{\perp} = 0$. But this contradicts TP, because there must be a 
2-valued homomorphism mapping $e_{r_{i}}$ onto 1, and this 
homomorphism would have to map both $p$ and $p^{\perp}$ onto 0. 

So if $r_{i}$ is included in $p$ or in $p^{\perp}$ then 
$e_{r_{i}} \leq p 
\mbox{ or } e_{r_{i}} \leq p^{\perp}$, and if $r_{i}$ is not included 
in $p$ or in $p^{\perp}$, then $e_{r_{i}}$ must belong to ${\cal 
D}(e,R)$, from which it follows that $e_{r_{i}} \leq p 
\mbox{ or } e_{r_{i}} \leq p^{\perp}$.

To see that $e_{r_{i}}$ must belong to ${\cal D}(e,R)$ if (i) $b = 
r_{i} \wedge p$ is orthogonal to $e_{r_{i}}$ in $r_{i}$, or (ii) $b$ 
is skew to $e_{r_{i}}$ in $r_{i}$, consider each of these cases in 
turn.

(i) Suppose $b$ is orthogonal to $e_{r_{i}}$ in $r_{i}$; that is, 
$b \leq e_{r_{i}}'$ (where the $'$ here denotes the orthogonal 
complement in $r_{i}$). If $r_{i}$ is 1-dimensional, $e_{r{_i}} = r{_i}$ 
and it follows immediately from 
R-PREF that $e_{r{_i}} \in {\cal D}(e,R)$. If $r_{i}$ is 
2-dimensional, $b = e_{r_{i}}'$ and 
$e_{r_{i}} = r \wedge b^{\perp}$, so $e_{r_{i}} \in {\cal D}(e,R)$ by 
lattice closure 
(that is, by TP), because both $r_{i}$ and $b^{\bot}$ belong to 
${\cal D}(e,R)$. If $r_{i}$ is more than 2-dimensional, consider 
all lattice automorphisms $U$ that are rotations about $e_{r_{i}}$ 
in $r_{i}$ and the identity in $r_{i}^{\perp}$. Such rotations 
preserve $e$ and $R$, because they preserve the eigenspaces of $R$ and 
the projections of $e$ onto the eigenspaces of $R$. It follows that 
$U(b) \in {\cal D}(e,R)$, by DEF. There are clearly sufficiently 
many rotations to generate a set of elements ${U(b)}$ whose span is 
$e_{r_{i}}'$. So $e_{r_{i}} \in {\cal D}(e,R)$, and hence 
$e_{r_{i}}\in {\cal D}(e,R)$ by lattice closure.

(ii) Suppose $b$ is skew to $e_{r_{i}}$ in $r_{i}$. Consider first 
the case that the eigenspace $r_{i}$ is more than 2-dimensional. In 
that case, we may suppose that $b$ is not a ray (or else we apply the 
following argument to $b'$). The subspace $b$ can therefore be 
represented as the span of a subspace $c$ orthogonal to $e_{r_{i}}$ 
in $r_{i}$, and the ray $d$, the projection of the ray $e_{r_{i}}$ 
onto $b$. Consider a lattice automorphism $U$ that is a reflection 
through the hyperplane $e_{r_{i}} \vee c$ in $r_{i}$ and the 
identity in $r_{i}^{\perp}$. (So $U$ preserves $e_{r_{i}}$ and $c$, 
but not $d$.) As in (i) above, $U(b) \in  {\cal D}(e,R)$ because $U$ 
preserves $e$ and $R$, and $b \wedge U(b) = c$, so $c \in {\cal 
D}(e,R)$. We can now consider rotations of $c$ about $e_{r_{i}}$ in 
$r_{i}$ as in (i) to show that $e_{r_{i}}\;' \in {\cal D}(e,R)$, and 
hence that $e_{r_{i}} \in {\cal D}(e,R)$.

The argument fails if $r_{i}$ is 2-dimensional. If $b$ is skew to 
$e_{r_{i}}$ in $r_{i}$ and $r_{i}$ is 2-dimensional, then $b$ must be 
a ray in $r_{i}$; that is, $c$ is the null subspace and $d = b$. The 
automorphism $U$ reduces to a reflection through $e_{r_{i}}$ in 
$r_{i}$. Suppose $b$ is not at a $45^{\circ}$ angle to $e_{r_{i}}$. 
The four rays $b, b', U(b), U(b)'$ would all have to belong to ${\cal 
D}(e,R)$ if $b$ belongs to ${\cal D}(e,R)$, but this contradicts TP. A 
2-valued homomorphism would have to map one of the rays $b$ or $b'$ 
onto 1, and one of the rays $U(b)$ or $U(b)'$ onto 1, and hence the 
intersection of these two rays---the null subspace---would also have to 
be mapped onto 1. So $b$ cannot be skew to $e_{r_{i}}$ in $r_{i}$, 
unless $b$ is at a $45^{\circ}$ angle to $e_{r_{i}}$. In this case, 
reflecting $b$ through $e_{r_{i}}$ yields $b'$, and there is no 
contradiction with TP.

So, if $r_{i}$ is 2-dimensional (and $r_{i}$ is not included in $p$ 
or $p^{\bot}$), we could conclude from this argument only that 
$e_{r_{i}} \leq p$ or $e_{r_{i}} \leq p^{\perp}$, or that $b \leq 
p$ or $b \leq p^{\perp}$, 
where b is a ray at a $45^{\circ}$ angle to $e_{r_{i}}$ 
in $r_{i}$. In the original formulation of the 
proof, the `weak separability' condition was introduced to 
exclude this one anomalous possibility, that the determinate 
sublattice ${\cal D}(e,R)$ might contain propositions $p$ such 
that $b \leq p$ or $b \leq p^{\bot}$, where $b$ is at a 
$45^{\circ}$ angle to $e_{r_{i}}$ in $r_{i}$.

To see that such a condition is not required, suppose $r_{i}$ is 2-dimensional. 
Consider a lattice automorphism $U$ that is the identity in 
$r_{i}^{\bot}$, preserves $e_{r_{i}}$ in $r_{i}$, and maps $b$ 
onto $U(b) \neq b$. For example, suppose we represent $e_{r_{i}}$ by 
the vector $\left( \begin{array}{c} 1 \\ 0 \end{array} \right)$
in $r_{i}$ and take the action of $U$ on $r_{i}$ as represented by 
the matrix $\left( \begin{array}{cc} i & 0 \\ 0 & 1 \end{array} \right)$. 
If $b$ is represented by a vector 
$\left( \begin{array}{c} b_{1} \\ b_{2} \end{array} \right)$
in $r_{i}$, then:
\[\left( \begin{array}{cc} i & 0 \\ 0 & 1 \end{array} \right) 
\left( \begin{array}{c} 1 \\ 0 \end{array} \right) = 
\left( \begin{array}{c} i \\ 0 \end{array} \right) = 
i\left( \begin{array}{c} 1 \\ 0 \end{array} \right) \]

\[\left( \begin{array}{cc} i & 0 \\ 0 & 1 \end{array} \right) 
\left( \begin{array}{c} b_{1} \\ b_{2} \end{array} \right) = 
\left( \begin{array}{c} ib_{1} \\ b_{2} \end{array} \right) \neq 
k\left( \begin{array}{c} b_{1} \\ b_{2} \end{array} \right)\] 
for any constant $k$. So $U$ maps $e_{r_{i}}$ onto the same ray but 
maps  $b$ onto a different ray, whether or not $b$ is at a 
$45^{\circ}$ angle to $e_{r_{i}}$. The four rays $b$, $b'$, $U(b)$, $U(b)'$ 
would all have to belong to ${\cal D}(e,R)$ if $b$ belongs to 
${\cal D}(e,R)$, which contradicts TP, by the argument above. So, if 
$r_{i}$ is 2-dimensional, $b$ cannot be skew to $e_{r_{i}}$ in $r_{i}$.
 
We have now established that if $p \in {\cal D}(e,R)$ then either 
$e_{r_{i}} \leq p$ or $e_{r_{i}} \leq p^{\bot}$, for all $i = 
1, \ldots, k$. The final stage of the proof involves showing that the 
set of \emph{all} such elements $\{p: e_{r_{i}} \leq p$ or 
$e_{r_{i}} \leq p^{\bot}, \mbox{for all }i = 1, \ldots, k\} = 
{\cal L}_{e_{r_{1}}e_{r_{2}} \ldots e_{r_{k}}}$ 
satisfies the three conditions TP, R-PREF, and DEF. Maximality then 
requires that 
${\cal D}(e,R) =
 {\cal L}_{e_{r_{1}}e_{r_{2}} \ldots e_{r_{k}}}$. (For details, see 
(Bub, 1997).)

\end{document}